\documentclass{article}
\usepackage{spconfa4,amsmath,graphicx}
\usepackage{cite}

\usepackage{tabularx,booktabs}
\usepackage{multirow}

\title{Suppressing Noise Disparity in Training Data \\ for Automatic Pathological Speech Detection}
\name{Mahdi Amiri$^{1,2}$, Ina Kodrasi$^1$ \thanks{This work was supported by the Swiss National Science Foundation project CRSII5\_202228 on ``Characterisation of motor speech disorders and processes''}}
\address{$^1$Idiap Research Institute, Switzerland\\
$^2$École Polytechnique Fédérale de Lausanne, Switzerland\\
{\tt \{mahdi.amiri,ina.kodrasi\}@idiap.ch}}
\begin{document}
%

\maketitle

\begin{abstract}
Although automatic pathological speech detection approaches show promising results when clean recordings are available, they are vulnerable to additive noise.
Recently it has been shown that databases commonly used to develop and evaluate such approaches are noisy, with the  noise characteristics between healthy and pathological recordings being different. 
Consequently, automatic approaches trained on these databases often learn to discriminate noise rather than speech pathology. 
This paper introduces a method to mitigate this noise disparity in training data. 
Using noise estimates from recordings from one group of speakers to augment recordings from the other group, the noise characteristics become consistent across all recordings. 
Experimental results demonstrate the efficacy of this approach in mitigating noise disparity in training data, thereby enabling automatic pathological speech detection to focus on pathology-discriminant cues rather than noise-discriminant ones.

\end{abstract}
\section{Introduction}
\label{sec:intro}
Early identification of pathological speech conditions such as dysarthria or apraxia of speech may serve as an early indicator of neurological disorders like Parkinson’s disease, highlighting the critical need for swift diagnosis of pathological speech.
Traditionally, speech and language pathologists rely on auditory-perceptual tests to identify pathological speech.
However, these tests are costly, subjective, and time-consuming. 
As a result, there is a growing interest in the research community to explore methods for automatically detecting pathological speech.
Previously, researchers predominantly relied on handcrafted acoustic features coupled with traditional machine learning methods to detect pathological speech~\cite{kadi2016fully, handcrafted_1, handcrafted_2, narendra_interspeech, handcrafted_3}.  
More recently, the success of deep learning algorithms in many fields has prompted the adoption of deep learning approaches for automatic pathological speech detection~\cite{baseline, deep_learning_based, w2v2_Pathological_ref, mel_ref_vae, mfcc_ref, train_noise_ref, lstm_patho_ref}. 
These approaches leverage a variety of speech representations, such as e.g., the short-time Fourier transform (STFT)~\cite{baseline}, Mel frequency cepstral coefficients~\cite{mfcc_ref, train_noise_ref}, Mel spectrograms~\cite{mel_ref_vae}, or self-supervised embeddings like wav2vec2~\cite{w2v2_Pathological_ref}, integrated with diverse architectures to achieve automatic pathological speech detection. 
For example, the approach proposed in~\cite{baseline} has shown promising performance using a convolutional neural network (CNN) to extract pathology-discriminant cues from STFT input representations. 
Similarly, in~\cite{w2v2_Pathological_ref}, a high performance is achieved using linear layers to extract pathology-discriminant cues from wav2vec2 embeddings.

Despite the reported success of automatic pathological speech detection approaches, state-of-the-art literature typically assumes that recordings from both healthy and pathological speakers are acquired in identical, noise-free environments using the same recording setup. 
However, obtaining high-quality, clean recordings from pathological speakers, such as e.g., by recording all speakers in an anechoic chamber, poses a challenge. 
In~\cite{train_noise_ref} it has been shown that widely used databases in the research community, such as UA-Speech~\cite{uaspeech} and TORGO~\cite{torgo}, are noisy, with the noise characteristics in healthy recordings being distinctly different from the noise characteristics in pathological recordings.
Given the disparity in noise characteristics between these speaker groups, classifiers trained on such data for pathological speech detection capture noise-discriminant cues rather than pathology-discriminant ones~\cite{train_noise_ref}. 
Consequently, the assessment of pathological speech detection approaches using such databases remains inconclusive.
 


Although the robustness to noise has been extensively investigated in many speech applications such as automatic speech recognition~\cite{robust_data2vec}, audio event detection~\cite{salamon2017deep}, or speaker identification~\cite{mohammadamini2022comprehensive}, the robustness to noise of pathological speech detection approaches has received significantly less attention, particularly for recordings with noise disparity among the two groups of speakers.
In~\cite{vasquez2015automatic}, it has been proposed to use a single-channel speech enhancement module prior to training a pathological speech detection approach. 
However, employing traditional single-channel speech enhancement on recordings with varying noise characteristics leads to signal distortions dependent on these characteristics.
Hence, instead of capturing pathology-discriminant cues, automatic pathological speech detection models trained on enhanced recordings will learn cues associated with the introduced distortions.

To address the challenge posed by noise disparity in training data for pathological speech detection approaches, one can either develop approaches that are robust to noise or suppress this noise disparity. In this paper, we propose a method to suppress the noise disparity in training data through noise augmentation. 
Given the available noisy recordings, we suggest using a voice activity detection (VAD) module~\cite{vad_ref} to extract an estimate of the noise present in each recording. Subsequently, the estimated noise signals from recordings of one speaker group are incorporated into the recordings of the other group, with appropriate scaling factors computed to maintain consistent signal-to-noise ratios (SNRs). 
This procedure ensures that the noise characteristics (in terms of the noise type and SNR) are similar between the two speaker groups. 
Therefore, with noise-discriminant cues suppressed in the data, automatic pathological speech detection approaches prioritize learning pathology-discriminant cues over noise-discrminant ones.

\section{Problem Statement and \\ Proposed Method}
\label{sec:statement}

We consider a noisy signal $y$ at time index $k$ given by
\begin{equation}
y_k = s_k + n_k,
\end{equation}
with $s$ denoting the clean speech signal and $n$ denoting the noise signal.
The power of the noisy signal $y$ can be computed as
\begin{equation}
\label{noise_energ}
  P_y = \frac{1}{K} \sum_k y^2_{k},
\end{equation}
with $K$ denoting the signal length.
 Further, we define the SNR of the noisy signal $y$ with respect to the noise $n$ as
\begin{equation}
\label{eq: snr}
{\text{SNR}_{y}^{n}} = 20 \log_{10} \frac{\sqrt{P_{s}}}{ \sqrt{ P_{n}}}, 
\end{equation}
with $P_{s}$ and $P_n$ being the clean speech and noise powers defined similarly to~(\ref{noise_energ}).
For conciseness, the time index $k$ is omitted in the remainder of this section unless explicitly required.


Without loss of generality, we consider one noisy utterance $y_{h}$ from the healthy group and one noisy utterance $y_{p}$ from the pathological group.
These utterances (of possibly different length) are given by
\begin{align}
 {y_{h}} = {s_{h}} + {n_{h}}, \; \; \; \;  {y_{p}} = {s_{p}} + {n_{p}},
  \label{equation:eq1}
\end{align}
with $s_{h}$ and $s_{p}$ denoting the respective clean signals and $n_{h}$ and $n_{p}$ denoting the respective noise signals.
Since $n_h \neq n_p$ and ${\text{SNR}_{y_h}^{n_h}} \neq {\text{SNR}_{y_p}^{n_p}}$, a pathological speech detection classifier trained using $y_h$ and $y_p$ can easily learn the noise differences between the two signals rather than pathology-discriminant cues in $s_h$ and $s_p$.
To suppress the noise disparity between the two utterances, we propose to use noise augmentation and add (an estimate of) the noise present in one utterance to the other utterance as
\begin{align}
\label{eq: augment}
 \hat{y}_h = \underbrace{s_h + n_h}_{y_h} + \alpha_h \hat{n}_p, \; \; \hat{y}_p = \underbrace{s_p + n_p}_{y_p} + \alpha_p \hat{n}_h,
\end{align}
with $\hat{n}_p$ an estimate of $n_p$, $\hat{n}_h$ an estimate of $n_h$, and the scalars $\alpha_h$ and $\alpha_p$ computed as described in the following. 
Although both noises are present in both signals in~(\ref{eq: augment}), there remains a noise disparity between $\hat{y}_h$ and $\hat{y}_p$ since their SNRs might still differ.
For $\hat{y}_h$ and $\hat{y}_{p}$ to contain similar noise characteristics such that a network trained on these utterances fails to learn noise-discriminant cues, it is required that
\begin{itemize}
    \item the SNR of both noisy signals with respect to each of the noises is the same, and
    \vspace{-8pt}
    \item the SNR of both noisy signals with respect to the total noise is the same.
\end{itemize}
More precisely, these conditions can be expressed as
\begin{align}
\label{eq: condition1}
\text{SNR}_{\hat{y}_h}^{n_h} & = \text{SNR}_{\hat{y}_p}^{\alpha_p \hat{n}_h} \\
\label{eq: condition2}
\text{SNR}_{\hat{y}_h}^{\alpha_h \hat{n}_p} & = \text{SNR}_{\hat{y}_p}^{n_p} \\
\label{eq: condition3}
\text{SNR}_{\hat{y}_h}^{{n}_h + \alpha_h\hat{n}_p} & = \text{SNR}_{\hat{y}_p}^{n_p + \alpha_p\hat{n}_h },
\end{align}
with the different SNRs computed similarly to~(\ref{eq: snr}).
The system of equations in~(\ref{eq: condition1})-(\ref{eq: condition3}) is an over-determined system.
However, if $\hat{n}_p$ and $\hat{n}_h$ are good estimates of $n_p$ and $n_h$, one can assume that
\begin{equation}
\label{eq: ep}
P_{n_h} = P_{\hat{n}_h}, \; \; P_{n_p} = P_{\hat{n}_p},
\end{equation}
with $P_{\{\cdot\}}$ denoting the power of the different noise signals defined similarly to~(\ref{noise_energ}).
If the assumption in~(\ref{eq: ep}) holds, scalars $\alpha_h$ and $\alpha_p$ that satisfy~(\ref{eq: condition1})-(\ref{eq: condition3}) can be computed using the power of clean speech signals $P_{s_h}$ and $P_{s_p}$ as
\begin{equation}
\label{eq: alpha_condition}
\alpha_h = \sqrt \frac{ {P_{s_h} }}{ {P_{s_p} }}, \; \; \alpha_p = \sqrt \frac{ {P_{s_p} }}{ { P_{s_h} }}.
\end{equation}
Using these scalars in~(\ref{eq: augment}) and using $\hat{y}_h$ and $\hat{y}_p$ instead of $y_h$ and $y_p$ to train a pathological speech detection system results in a system that learns pathology-discriminant cues rather than noise-discriminant ones.

The proposed approach requires noise estimates $\hat{n}_p$ and $\hat{n}_h$~(cf.~(\ref{eq: augment})) as well as estimates of the clean speech powers $P_{s_p}$ and $P_{s_h}$~(cf.~(\ref{eq: alpha_condition})).
To obtain these estimates, we propose to extract the noise-only segments from each utterance using a VAD and assume that they yield a good approximation of the true noise.
Such an assumption holds for relatively stationary noise, which is a reasonable assumption in clinical settings where these recordings are typically collected.
The extracted noise-only segments are then concatenated and repeated as necessary to obtain noise signals $\hat{n}_p$ and $\hat{n}_h$ of appropriate lengths required in~(\ref{eq: augment}).
Furthermore, assuming that the speech and noise signals are uncorrelated, the clean speech powers are computed as
\begin{equation}
\label{eq: cs}
P_{s_h} = P_{y_h} - P_{n_h},  \; \; P_{s_p} = P_{y_p} - P_{n_p},
\end{equation}
with $P_{n_h}$ and $P_{n_p}$ computed using the extracted noise-only segments.
Although we have made several assumptions and approximations which may not hold in practice, experimental results in Section~\ref{sec:results} show that these assumptions and approximations are effective in practice to (partially) suppress the noise disparity in the training data.

\begin{table*}[!ht]
\caption{Mean and standard deviation of the classification accuracy (\%) of the  CNN-based and wav2vec2-based approaches using different training procedures for different SNR settings.
}
\label{tab: results}
\begin{center}
\begin{small}
\begin{sc}
\begin{tabularx}{\linewidth}{X|ccccccc}
\toprule

\multirow{2}{*}{SNR Setting} &\multirow{2}{*}{Test} &\multicolumn{3}{|c}{CNN-based} & \multicolumn{3}{|c}{wav2vec2-based} \\

&&\multicolumn{1}{|c}{Standard} & \multicolumn{1}{|c}{Oracle} &  \multicolumn{1}{|c}{Practical}& \multicolumn{1}{|c}{Standard} & \multicolumn{1}{|c}{Oracle} & \multicolumn{1}{|c}{Practical} \\

\midrule

\multirow{2}{*}{A} & \multirow{1}{*}{Noisy} 
& \multicolumn{1}{|c}{$99.9 \pm 0.1$} & \multicolumn{1}{|c}{$70.5 \pm 4.8$} & \multicolumn{1}{|c}{$76.3 \pm 3.2$} & \multicolumn{1}{|c}{$98.1 \pm 0.6$} & \multicolumn{1}{|c}{$78.0 \pm 1.7$} & \multicolumn{1}{|c}{$86.3 \pm 2.7$} \\

& \multirow{1}{*}{Clean}

& \multicolumn{1}{|c}{$54.1 \pm 5.8$} & \multicolumn{1}{|c}{$68.3 \pm 1.7$} & \multicolumn{1}{|c}{$67.4 \pm 2.0$} & \multicolumn{1}{|c}{$70.3 \pm 4.6$} & \multicolumn{1}{|c}{$77.5 \pm 1.6$} & \multicolumn{1}{|c}{$77.3 \pm 1.2$} \\ \hline

\multirow{2}{*}{B} & \multirow{1}{*}{Noisy} 
& \multicolumn{1}{|c}{$84.8 \pm 1.7$} & \multicolumn{1}{|c}{$75.9 \pm 1.3$} & \multicolumn{1}{|c}{$70.8 \pm 0.8$} & \multicolumn{1}{|c}{$82.9 \pm 1.6$} & \multicolumn{1}{|c}{$81.1 \pm 0.9$} & \multicolumn{1}{|c}{$82.5 \pm 0.2$} \\

& \multirow{1}{*}{Clean}

& \multicolumn{1}{|c}{$77.2 \pm 2.3$} & \multicolumn{1}{|c}{$74.6 \pm 1.1$} & \multicolumn{1}{|c}{$69.7 \pm 0.4$} & \multicolumn{1}{|c}{$81.5 \pm 1.6$} & \multicolumn{1}{|c}{$81.2 \pm 0.3$} & \multicolumn{1}{|c}{$82.4 \pm 0.6$} \\ \hline

\multirow{2}{*}{C} & \multirow{1}{*}{Noisy} & 
\multicolumn{1}{|c}{$99.5 \pm 0.5$} & \multicolumn{1}{|c}{$71.5 \pm 6.5$} & \multicolumn{1}{|c}{$76.7 \pm 3.8$} & \multicolumn{1}{|c}{$95.5 \pm 2.8$} & \multicolumn{1}{|c}{$78.5 \pm 2.0$} & \multicolumn{1}{|c}{$86.2 \pm 3.0$} \\

& \multirow{1}{*}{Clean}

& 
\multicolumn{1}{|c}{$51.3 \pm 1.9$} & \multicolumn{1}{|c}{$70.3 \pm 1.9 $} & \multicolumn{1}{|c}{$68.2 \pm 1.7$} & \multicolumn{1}{|c}{$53.6 \pm 5.9$} & \multicolumn{1}{|c}{$80.0 \pm 2.1$} & \multicolumn{1}{|c}{$77.6 \pm 5.0$} \\

\bottomrule
\end{tabularx}
\end{sc}
\end{small}
\end{center}
\vspace{-.2cm}
\end{table*}

\section{Experimental Settings}
For the experimental results, we adopt a methodology similar to that of~\cite{train_noise_ref}.
We consider noisy healthy and pathological speech recordings, with the noise characteristics differing among the two groups.
Further, we consider two state-of-the-art automatic pathological speech detection approaches, i.e., the CNN-based approach from~\cite{baseline,amiri_eusipco_2024} and the wav2vec2-based approach from~\cite{w2v2_Pathological_ref,amiri_eusipco_2024}.
 To evaluate the impact of noise disparity on automatic pathological speech detection approaches, these approaches are trained using noisy utterances and their performance on noisy and clean test utterances is compared.
To evaluate the impact of suppressing the noise disparity using the proposed method, these approaches are trained using augmented utterances and their performance on noisy and clean test utterances is compared.
Obtaining a better classification performance on noisy test utterances than on clean test utterances confirms that noise-discriminant cues are being learnt instead of pathology-discriminant cues.
Obtaining a similar classification performance on both noisy and clean test utterances confirms that pathology-discriminant cues are being learnt instead of noise-discriminant cues.

In the following, we describe the experimental settings used in our evaluation.
\label{sec:setting}
\subsection{Databases}
Since we do not have access to the clean speech signals in the UA-Speech and TORGO databases, we construct synthetic noisy databases for our analysis.

\label{clean_data}
{\emph{Clean speech.}} \enspace We consider clean recordings of healthy and pathological speakers from the PC-GITA database~\cite{pc_gita}. 
This database contains Spanish recordings from gender-balanced groups of $50$ healthy speakers and $50$ patients suffering from Parkinson's disease.
Recordings are downsampled to $16$~kHz prior to using them for our experiments.

{\emph{Noise.}} \enspace
To generate noisy recordings, we augment the clean recordings with different noise types from the DEMAND database \cite{demand}. 
Three different noise types are used, i.e., D-type (DKITCHEN and DLIVING), N-type (NPARK and NRIVER), and O-type (OOFFICE and OMEETING).
To simulate noise disparity in the healthy and pathological recordings, one group of speakers is augmented with one specific noise from a given type (e.g., DKITCHEN used for healthy speakers), whereas the other group of speakers is augmented with the other noise from the same type (e.g., DLIVING used for pathological speakers).
For each of the three considered noise types, we generate data for three different SNR settings, i.e., 
\vspace{-3pt}
\begin{itemize}
\item[A.] healthy and pathological recordings have the same SNR of $20$~dB,
\vspace{-8pt}
\item[B.] healthy and pathological recordings have the same SNR of $40$~dB, and
\vspace{-8pt}
\item[C.] healthy recordings have an SNR of $20$~dB whereas pathological recordings have an SNR of $40$~dB.
\vspace{-3pt}
\end{itemize}
The choice of such relatively high SNR values is done to reflect the (more recent) reality where care is taken to have better quality pathological recordings as well as to show that even an SNR of $40$~dB is problematic for state-of-the-art pathological speech detection approaches.


\subsection{Model Architecture}
\label{architecture}




\emph{CNN-based approach.} \enspace 
The CNN-based approach operates on fixed-size segments of speech.
To compute inputs to this approach, we segment utterances into $500$~ms long segments (with an overlap of $250$~ms) and compute their STFT. 
The used STFT parameters and the architecture of the CNN-based approach is the same as in~\cite{amiri_eusipco_2024}.


\emph{wav2vec2-based approach.} \enspace
Differently from the CNN-based approach, the wav2vec2-based approach operates on full utterances of variable length.
As in~\cite{amiri_eusipco_2024}, the wav2vec2-base model is frozen and used to extract features from full utterances.
These features are then used as input to a linear layer (input size $768$, output size $256$).
After a ReLU activation function, batch
normalization, and dropout ($p = 0.3$), a final linear layer (input size $256$, output size $2$) is used for pathological speech detection.


\subsection{Training and Evaluation}
\label{training}
A stratified $10$-fold cross validation framework is used to evaluate the performance of the proposed method, with no overlap between speakers in the training, validation, and test sets.

As in~\cite{amiri_eusipco_2024}, the CNN-based approach is trained using the SGD optimizer with a learning rate of $0.001$, whereas the wav2vec2-based approach is trained using the Adam optimizer with a learning rate of $0.01$. 
For both approaches, we use a weight decay of $5 \times 10^{-4}$ and the learning rate scheduler \textit{ReduceLROnPlateau} with $patience= 5$ and $factor= 0.5$.
Training is stopped if the learning rate decreases beyond $10^{-4}$ of the initial learning rate or if the maximum number of epochs of $100$ is reached.

To implement the method proposed in~Section~\ref{sec:statement}, noise is added to each utterance from each group of speakers according to~(\ref{eq: augment}).
This noise is extracted from a randomly selected utterance from the other group of speakers.
This procedure is repeated in each epoch.

The performance is evaluated using the speaker-level accuracy computed similarly as in~\cite{amiri_eusipco_2024}. 
To account for the effect of randomness in initializing networks, we analyze the mean performance for networks trained with $5$ different seeds. 

\section{Experimental Results}
\label{sec:results}

In this section, we present several results to validate the effectiveness of the proposed method.
To demonstrate the validity of the proposed method and to decouple it from potential estimation errors in practice, we consider the oracle scenario where one has access to the noise signals required for augmentation in~(\ref{eq: augment}) and to the clean speech powers required for computing appropriate scalars in~(\ref{eq: alpha_condition}).
To demonstrate the effectiveness of the proposed method in practice, we also consider the practical scenario where the different required quantities are estimated through the VAD as described in Section~\ref{sec:statement}.
The performance obtained when using the proposed method in these scenarios is compared to the performance when the standard training procedure is used, i.e., without accounting for the existing noise disparity in the training data.
All results are given in Table~\ref{tab: results}, where the mean and standard deviation of the performance across the different considered noise types is presented.


\emph{Standard training.} \enspace
Table~\ref{tab: results} shows that in all SNR settings, using standard training for the CNN-based approach yields a very high performance on noisy test utterances and 
a considerably lower performance on clean test utterances. Although the wav2vec2-based approach is expected to be more robust to noise, using standard training for this approach also results in a considerably higher performance on noisy test utterances than on clean test utterances for SNR settings A and C.
These results confirm that in the presence of noise disparity in the data, these approaches learn noise-discriminant cues rather than pathology-discriminant cues.

\emph{Oracle scenario.} \enspace
Table~\ref{tab: results} shows that when using the proposed method with oracle quantities, the performance of both approaches on noisy and clean test utterances is very similar for all SNR settings.
These results confirm the validity of the proposed method, with noise augmentation being effective at suppressing noise disparity in the data and allowing pathological speech detection approaches to learn pathology-discriminant cues instead of noise-discriminant cues.

\emph{Practical scenario.} \enspace
Table~\ref{tab: results} shows that in comparison to the standard training procedure, the proposed method with realistically estimated quantities increases the performance on clean test utterances and decreases the performance on noisy test utterances as desired.
Nevertheless, there remains a gap between the effectiveness of the proposed method in practice in comparison to the oracle implementation for SNR settings A and C. 
Additional investigations confirm that the estimated clean speech powers are very similar to the oracle speech powers, and hence, this remaining performance gap can be attributed to the approximation of the noise by concatenating and repeating noise-only segments extracted through the VAD.
In the future, alternative methods to extract the noise-only signals will be investigated.

\section{Conclusion}
\label{sec: conclusion}
In this paper we have proposed a method to suppress the noise disparity in training data for automatic pathological speech detection. 
The goal is to enable these approaches to prioritize learning pathology-discriminant cues over noise-discriminant ones. 
The proposed method involves extracting noise from recordings using a VAD and then augmenting recordings from one speaker group with noise extracted from another speaker group. Extensive experimental results have demonstrated the effectiveness of the proposed method.


\small
\bibliographystyle{IEEEtran}
\bibliography{refs}

\end{document}